\documentclass[aps,prd,showkeys,showpacs,twocolumn,superscriptaddress]{revtex4}

\usepackage[dvips]{graphicx}
\usepackage{amsfonts}
\usepackage{amsmath}
\usepackage{latexsym}
\usepackage{amssymb}
\usepackage[mathcal]{euscript}

\newcommand{\ket}[1]{| #1 \rangle}
\newcommand{\bra}[1]{\langle #1 |}

\newcommand{\ul}{\underline}
\newcommand{\ep}{\varepsilon}
\newcommand{\epm}{\ul{\ep_m}}
\newcommand{\vo}{\ul{v^0}}
\newcommand{\vi}{\ul{v^1}}
\newcommand{\vx}{\ul{v^x}}
\newcommand{\trace}{\textrm{Tr}}
\newcommand{\E}{\sqrt{\varepsilon}}

\newtheorem{Lem}{Lemma} 
\newtheorem{Prop}{Proposition}
\newtheorem{Cor}{Corollary}

\begin{document}

\title{\begin{center}
The Conal representation of Quantum States
\\ and Non Trace-Preserving Quantum Operations
\end{center}
}

\author{Pablo Arrighi}
\email{pja35@cl.cam.ac.uk}
\affiliation{Computer Laboratory, University of Cambridge, 15 JJ
Thomson Avenue, Cambridge CB3 0FD, U.K. }
\author{Christophe Patricot \setcounter{footnote}{5}}
\email{cep29@damtp.cam.ac.uk}
\affiliation{
DAMTP, University of
Cambridge,
Centre for Mathematical Sciences,\\ Wilberforce Road, Cambridge CB3 0WA,
U.K.}



\begin{abstract}
We represent generalized density
matrices \vspace{-0.7 mm}of a $d$-complex dimensional quantum system as a subcone of a
real pointed cone of revolution in $\mathbb{R}^{d^2}$, or indeed a
Minkowskian \vspace{-0.7 mm}cone in $\mathbb{E}^{1,d^2-1}$. Generalized pure states
correspond to certain future-directed light-like vectors of
$\mathbb{E}^{1,d^2-1}$. This extension of the Generalized Bloch Sphere
enables us to cater for non-trace-preserving quantum operations, and
in 
particluar to view the per-outcome effects of generalized
measurements. We show that these consist of the product of an orthogonal
transform 
about the axis of the cone of revolution and a positive real
linear transform. We give detailed formulae for the one qubit case and 
express the post-measurement states in terms of the initial state
vectors and
measurement vectors. We 
apply these results in order to find the information gain versus
disturbance tradeoff in the case of two equiprobable pure states. Thus
we recover
Fuchs and Peres' formula in an elegant manner.  
\end{abstract}

\maketitle

\section{Introduction}

The space of pure states of finite $d$-dimensional Quantum Mechanics
$\mathbb{C}P^d$, set of rays in the complex Hilbert space
$\mathbb{C}^d$, is, as most complex spaces, not easy to visualize.
Physical motions, let alone unitary time evolutions, have no clear
geometric interpretation. However, the set of hermitian operators
on $\mathbb{C}^d$, $\textrm{Herm}_d(\mathbb{C})$, is a
$(d^2\!-1)$-dimensional \emph{real} vector space, and as such is
certainly easier to represent geometrically. States, or more
generally density matrices, form of course a subset of
$\textrm{Herm}_d(\mathbb{C})$. The Generalized Bloch Sphere
Representation (\cite{Eberly}-\cite{Mahler}) is a famous application of this fact which has
proved to be popular and elucidating: a given density matrix can
be represented as a real vector inside a (hyper) sphere. \\
It turns out that this representation defined for density matrices, or
unit trace
positive elements of $\textrm{Herm}_d(\mathbb{C})$,
is only good at handling unitary or trace preserving quantum
operations on density matrices: the former induce rotations of the
Bloch vector, the latter affine transformations \cite{Zanardi}. Individual
outcomes of generalized measurements, for example, are not
directly representable. Considering the insight the Bloch Sphere
representation gave to unitary and trace-preserving operations, it
seems interesting, for the mere sake of geometry at first, but
mainly to give a useful picture to tackle Quantum Information
problems,  to extend it to cater for non-trace-preserving quantum
operations. This is further motivated by the fact that the space of
(semi-definite or definite) positive hermitian operators, thereafter denoted
$\textrm{Herm}_d^+(\mathbb{C})$,
\emph{is} a closed convex cone, and that all admissible quantum
operations should be a subset of the transformations of this
cone.

In spite of being so central in Quantum Information Theory, the
tradeoff between how much Shannon Information one may gain about a
quantum system \emph{versus} how much Disturbance the observation must
necessarily cause to the system, remains
extremely difficult to quantify. Quantum cryptographists tend to
circumvene the problem: most of their proofs are a witty blend of
the particular symmetries of the protocol in question, together
with a convoluted machinery. A few attempts have been made to
solve the tradeoff \cite{Banaszek}\cite{Barnum}, but only one \cite{Fuchs} deals with discrete
ensembles - namely the case of two equiprobable pure states, and
this is already something. Unfortunately the approach involves
lengthy algebra and a number of assumptions. One should
be able to find a method which gives a glimpse of intuition about
the geometry of optimal measurements, and for this purpose, we think
that our approach is useful. 

In section \ref{part one} we consider general quantum systems of $d$
complex dimensions. We give a representation of the set of positive
hermitian matrices $\textrm{Herm}_d^+(\mathbb{C})$ as a subcone of a real
Minkowskian cone in $\mathbb{R}^{d^2}$, and analyse
geometrical properties of generalized measurements in this setting. We
find that our approach is particularly useful to represent per-outcome
post-measurement states, and that pure states correspond to certain
light-like vectors of the cone. Unitary operators on the complex
system become real orthogonal transforms, while positive operators
become real positive transforms. Section \ref{part two} should be of
special interest for quantum information theorists: we treat the $d=2$ one qubit
case in full detail. We find further geometrical relations between
measurement vectors, state vectors and post-measurement state vectors
and give explicit formulae. In section \ref{part three} we apply our
results to a typical Information gain versus Disturbance tradeoff scenario in which Alice gives Eve two
equiprobable pure states. Thus we recover Fuchs and Peres' formula in an elegant
and geometrical manner.

\section{Conal representation of $d$-dimensional quantum systems}
\label{part one}
The state of such a system is described by a $d\times d$ density matrix. We
shall express hermitian matrices 
as real linear combinations of Hilbert-Schmidt-orthogonal hermitian
matrices, and then restrict this representation to elements of
$\textrm{Herm}_d^+(\mathbb{C})$, or generalized density
matrices. $\textrm{Herm}_d^+(\mathbb{C})$ turns out to be
``isomorphic'' to a convex subcone of a cone of revolution in
$\mathbb{R}^{d^2}$, or indeed a Minkowskian future cone in
$\mathbb{E}^{1,d^2-1}$. We then analyse the effects of quantum
operations on density matrices in this representation.

\subsection{Hermitian matrices}
Let $\{\tau_i\}$, $i\in\{1,\ldots,d^2-1\}$, be a Hilbert-Schmidt
orthogonal basis (as in (\ref{hilbert-schmidt})) of
$d \times d$ traceless hermitian matrices, and let $\tau_0$ be the
identity matrix $\mathbb{I}$. Throughout this article latin indices will run from
$1$ to $d^2-1$, greek indices from $0$ to $d^2-1$, and repeated indices are summed unless specified. We take the $\tau_{\mu}$'s to satisfy by
definition:
\begin{equation}
\label{hilbert-schmidt}
\forall \;\mu,\nu \quad \textrm{Tr}(\tau_{\mu}\tau_{\nu})=
d\delta_{\mu\nu}
\end{equation}
with $\delta$ the Kronecker delta. $\{\tau_{\mu}\}_{\mu}$ is a basis of
$\textrm{Herm}_d(\mathbb{C})$, and any
hermitian matrix $A\in \textrm{Herm}_d(\mathbb{C})$ decomposes on this
basis as 
\begin{align}
A&=\frac{1}{d}\big(\trace(A)\mathbb{I}+\trace(A\tau_i)\tau_i\big)
\nonumber \\
\label{components}
 &=\frac{1}{d}\trace(A\tau_{\mu})\tau_{\mu}
\end{align}
Letting $\ul{A}=(\ul{A}_{\mu})\in \mathbb{R}^{d^2}$ with
$\ul{A}_{\mu}=\trace(A\tau_{\mu})$ be the component vector of $A$ in
this particular basis, we have  
\begin{align}
&\forall \;A,B\in \textrm{Herm}_d(\mathbb{C}), \quad
AB=\frac{1}{d^2}\ul{A}_{\mu}\ul{B}_{\nu}\tau_{\mu}\tau_{\nu} \nonumber
\\
\label{isometry}
&\textrm{hence} \quad \trace{AB}=\frac{1}{d}\ul{A}.\ul{B}\equiv \frac{1}{d}\ul{A}_{\mu}\ul{B}_{\mu} 
\end{align}
We shall call $\ul A$ the vector in $\mathbb{R}^{d^2}$,
$\overrightarrow{A}=(\ul A_i)$ the restricted vector in
$\mathbb{R}^{d^2-1}$, and $\phi$ the coordinate map: 
\begin{align*}
\phi: \textrm{Herm}_d(\mathbb{C})&\to \mathbb{R}^{d^2} \\ 
        A & \mapsto \ul A
\end{align*}
Equation (\ref{isometry}) says that $\phi$ is an isometric isomorphism of
$(\textrm{Herm}_d(\mathbb{C}), \trace(\;))$ onto $(\mathbb{R}^{d^2},
(1/d)(\,.\,))$. Therefore any linear operator $L$ on $\textrm{Herm}_d(\mathbb{C})$
defines via $\phi$ and $\phi^{-1}$ an operator on $\mathbb{R}^{d^2}$, $M(L)=\phi\circ L \circ \phi^{-1}$. This definition
yields the following ``morphism'' property :
\begin{Lem}
If $L_1, L_2$ are linear operators on $\textrm{Herm}_d(\mathbb{C})$, then
$M(L_i)=\phi \circ L_i \circ \phi^{-1}$ for $i\!=1,2$ are
endomorphisms of 
$\mathbb{R}^{d^2}$ and satisfy 
\begin{equation}
M(L_1\circ L_2)=M(L_1)M(L_2)
\end{equation}
\end{Lem} 
In particular, any complex $d\times d$ matrix $A$ defines via
$Ad_A:\rho\mapsto A\rho A^{\dagger}$ a linear operator on
$\textrm{Herm}_d(\mathbb{C})$ which  corresponds to a real endomorphism $M(Ad_A):\ul \rho
\mapsto M(Ad_A)\ul \rho$ ; and $Ad_{AB}=Ad_A\circ Ad_B$ implies
$M(Ad_{AB})=M(Ad_A)M(Ad_B)$. As a direct consequence of this  and the previous
definitions , calling $GL_n(\mathbb{K})$ the group of invertible
$n\times n$ matrices on the field $\mathbb{K}$, 
we get :
\begin{Lem}
\label{subgroup}
For any subgroup $G$ of $GL_d(\mathbb{C})$, the following mapping
\begin{align}
\psi :G &\to \psi(G) \subset GL_{d^2}(\mathbb{R}) \nonumber \\
     A&\mapsto M(Ad_A)=\phi^{-1}\circ Ad_A \circ \phi  
\end{align}
is a group homomorphism. $\psi(G)$ is a subgroup of $GL_{d^2}(\mathbb{R})$. 
\end{Lem}
Note that since $\psi(\mathbb{I})=\psi(-\mathbb{I})= \mathbb{I}$,
$\psi$ is not necesseraly injective. Moreover $\psi$ is certainly not linear.  An interesting subgroup is the
Special Unitary group $SU(d)=\{U \in
GL_d(\mathbb{C})\;/\;UU^{\dagger}=\mathbb{I},\;
\textrm{det}\,U=1\}$. We call $SO(n)=\{O\in
GL_n(\mathbb{R})\;/\;OO^T=\mathbb{I}, \; \textrm{det}\;O=1\}$ the
special orthogonal group in $n$-dimensions.
\begin{Lem}
\label{unitary}
Special Unitary transformations on
$\textrm{Herm}_d(\mathbb{C})$, $Ad_U : \rho\mapsto U\rho U^{\dagger}$
with $U\in SU(d)$, induce rotations of
$\mathbb{R}^{d^2}$ about the $\mathbb{I}$-axis. In fact the linear transforms   
$\psi(U): \ul \rho \mapsto
\psi(U) \ul \rho$  are special orthogonal and $\psi(SU(d))$ is a subgroup
of $SO(d^2-1)$. It  is a proper
subgroup when $d\geq3$. Moreover, $\psi(U(d))=\psi(SU(d))$.
\end{Lem}
\textbf{Proof:} Let $\rho=(1/d)(\trace(\rho)\mathbb{I}+\ul
\rho_i\tau_i)$ a hermitian matrix. Using (\ref{isometry}) and the fact that $Ad_U$ is
trace-preserving for $U$ unitary:
\begin{align*}
\ul{U\rho U^{\dagger}}.\ul{U\rho
U^{\dagger}}&=(\trace\rho)^2+\big(\psi(U)\ul \rho \big)_i
\big(\psi(U)\ul \rho\big)_i \\
&= d\trace(U\rho U^{\dagger}U\rho U^{\dagger}) \\
&=d\trace{\rho^2}= \ul \rho .\ul \rho \\
&= (\trace\rho)^2 + \ul \rho_i\,\ul \rho_i 
\end{align*}
In addition to preserving the first component $\ul \rho_0=\trace \rho$,
$\psi(U)$ preserves the $\mathbb{R}^{d^2-1}$ scalar product $\ul{
\rho}_i.\ul{\rho}_i$. For all $U \in SU(d)$, there exist $t\in\mathbb{R}$
and $B\in su(d)$ such
that $U=U(t)=\textrm{exp}(tB)$.
Since $\textrm{det}\,\psi(U(0))=1$ 
and $t\mapsto \textrm{det}\,\psi(U(t))$ is
continuous and has values in $\{\pm1\}$,
$\textrm{det}\,\psi(U)=1$. Thus $\psi(U)$ is a special rotation
about the $\mathbb{I}$ -axis of
$\mathbb{R}^{d^2}$. By Lemma \ref{subgroup},
$\psi(SU(d))$ is a subgroup of the special orthogonal group
$SO(d^2-1)\subset
SO(d^2)$. As for all $\theta\in \mathbb{R}$, $Ad_U=Ad_{e^{i\theta}U}$,
$\psi(U(d))=\psi(SU(d))$. \\
Since the $\{\sqrt{-1}\tau_i\}$ span the Lie algebra
$su(d)$, $\psi(SU(d))$ is the Adjoint group of $SU(d)$. For $d=2$, we
get the whole of $SO(3)$, but this is not the case for $d>2$, as is easily seen looking at the dimensions:
\begin{align*}
\textrm{dim}\,SU(d)&=d^2-1 \\
\textrm{dim}\,SO(d^2-1)&=\frac{1}{2}(d^2-1)(d^2-2) \\
\end{align*}
and $\textrm{dim}\,SU(d)<\textrm{dim}\,SO(d^2-1)$ for $d>2$.$\quad \Box$\\
All the results of this section remain true of course when we just
consider $\textrm{Herm}^+_d(\mathbb{C})$. From now on, for any $A$ complex $d\times
d$ matrix we shall denote $\psi(A)=M(Ad_A)$ the real endomorphism of $\mathbb{R}^{d^2}$. 
\subsection{Generalized density matrices}
In \cite{Zanardi}, Zanardi showed using a restriction of a mapping
analogous to $\phi:A\mapsto \ul A$ that $d\times d$ density matrices
lie in a convex subset of a ball $S\subset \mathbb{R}^{d^2-1}$. We shall extend this
to a convex cone
by considering generalized density matrices, by which we mean elements of
$\textrm{Herm}^+_d(\mathbb{C})$. This seems more natural in the sense
that we like to think of the space of states of most physical theories
and indeed Quantum Mechanics   
as a space invariant under positive linear combinations and not just
convex combinations. In addition,
this bigger space
allows a per-outcome representation of generalized measurements. \\
We define generalized pure states to be generalized density
matrices which yield pure states after rescaling them to unit
trace. Note that these are not the ``states of partial purity'' of the
complex $d$-dimensional system, which are singular
density matrices. In other words, generalized pure states are not the
elements of the boundary of $\textrm{Herm}^+_d(\mathbb{C})$ in the
sense of characteristic functions of cones (see \cite{BPS} for example).  
\begin{Prop}
\label{cone}
The cone of positive hermitian matrices
$\textrm{Herm}^+_d(\mathbb{C})$ is isomorphic to a convex subcone
$C$ of the following cone of revolution in $\mathbb{R}^{d^2}$:
\begin{equation}
\label{light-cone}
\Gamma= \{(\lambda_{\mu})\in
\mathbb{R}^{d^2} / \sum_{i=1}^{d^2-1}\lambda_i^2 \leq
(d-1)\lambda_0^2, \lambda_0 \geq 0 \}
\end{equation} 
The set of generalized pure states verifies $\mathcal{C}=C\cap\partial
\Gamma$, where $\partial\Gamma$ stands for the boundary of $\Gamma$.
\end{Prop}
$\mathbf{Proof:}$ We begin as in \cite{Zanardi}. Let $\mathcal{P}$ denote the space of (not
generalized) pure states in
$\textrm{Herm}_d(\mathbb{C})$. In addition to being positive,
$\rho\in\mathcal{P}$ satisfies 
$\trace(\rho^2)=\trace(\rho)=1$, so we have 
\begin{align}
\trace(\rho^2)&=\frac{1}{d}\ul \rho.\ul
\rho=\frac{1}{d}\big((\trace\rho)^2+ \ul \rho_i\ul \rho_i \big)
\nonumber \\
 &=\frac{1}{d}\big(1+ \ul \rho_i\ul \rho_i \big)=1, \quad
\textrm{hence} \nonumber \\
\label{sphere}
\ul \rho_i\ul \rho_i&= d-1
\end{align}
The restricted vector $(\ul \rho_i)$ is on a $(d^2-2)$-sphere of
radius $\sqrt{d-1}$, $\partial S^{d^2-2}$, where $S$ is the corresponding
ball. In $\mathbb{R}^{d^2}$, $\ul \rho$ pure sits in the
intersection of the cylinder (\ref{sphere}) and the $\ul
\rho_0=\trace(\rho)=1$ hyperplane, in other words on $\partial S^{d^2-2}$ ``centered'' at
$(1,0,\ldots,0)$. \\
Any density matrix can be expressed as a positive (convex) linear combination
of pure states, and any positive (convex) linear combination of pure
states defines a density matrix. Calling $D$ the set of (not
generalized) density
matrices, $D\subset\overline{\textrm{Hull}(\mathcal{P})}$ and 
$\textrm{Hull}(\mathcal{P})\subset D$. Since $D$ is closed,
$D=\overline{\textrm{Hull}(\mathcal{P})}$, a well-known result. As
$\phi:\textrm{Herm}_d(\mathbb{C})\to \mathbb{R}^{d^2}$ is linear and
bi-continuous, $\phi(D)=\phi\big(\overline{\textrm{Hull}(\mathcal{P})}\big)=
\overline{\phi(\textrm{Hull}(\mathcal{P}))}=\overline{\textrm{Hull}(\phi(\mathcal{P}))}$.
This set is a closed
convex subset of  $S$ ``centered'' at $(1,0,\ldots,0)$:
\begin{displaymath}
\phi(\mathcal{P})\subset \partial S^{d^2-2} \Rightarrow
\overline{\textrm{Hull}(\phi(\mathcal{P}))}\subset S
\end{displaymath}
Calling $S^+\equiv\overline{\textrm{Hull}(\phi(\mathcal{P}))}$ the image
set of density matrices as a subset of $\mathbb{R}^{d^2-1}$, we get:
\begin{displaymath}
\rho\in D \Leftrightarrow \trace(\rho)=\ul \rho_0=1 \;\textrm{and} \; (\ul
\rho_i) \in S^+
\end{displaymath}
Now a non-zero $\ul \rho \in \textrm{Herm}_d(\mathbb{C})$ is
positive if and only if $(1/\trace\rho)\rho$ is positive, that is if
and only if $\big((1/\trace(\rho))\ul \rho_i\big) \in S^+$. In
$\mathbb{R}^{d^2}$, recalling that $\trace(\rho)\equiv \ul \rho_0$, this reads 
\begin{equation}
\rho \in \textrm{Herm}^+_d(\mathbb{C}) \Leftrightarrow \ul \rho \in
\{(\lambda_0, (\lambda_i))\in \mathbb{R}^{d^2} / (\lambda_i)\in
\lambda_0 \,S^+ \}
\end{equation}
This clearly defines a cone $C$ in $\mathbb{R}^{d^2}$. As
$S^+\subset S$, $C$ is a subcone of
the cone of revolution $\Gamma$ given by (\ref{light-cone}). $\phi$ being an isomorphism,
$C$ is convex and isomorphic to $\textrm{Herm}^+_d(\mathbb{C})$. As
pure states correspond to some points on the sphere $\partial S^{d^2-2}$,
generalized pure states lie  in the boundary of $\Gamma$. Calling $\mathcal{C}$ the set of vectors of $C$ corresponding to generalized
pure states, we have $\mathcal{C}\subset C\cap \partial
\Gamma$. Moreover $\mathcal{C}\supset C\cap \partial \Gamma$ follows from the fact that any rescaled positive matrix
$\rho$ such that $\trace(\rho^2)=\trace\rho= 1$ is a pure state. Remember that $\mathcal{C}$ is not the boundary of $C$, but a
cone over $\phi(\mathcal{P})$, the image set of pure states.  $\Box$ \\

As we shall
see in detail in section \ref{part two} in $d=2$ dimensions, generalized pure states correspond to
future-directed light-like vectors of Minkowski space of signature
$(1,3)$. We have shown that this
remains true to a certain extent in $d$-complex dimensions, $\Gamma$
being the future light-cone of Minkowski space $\mathbb{E}^{1,d^2-1}$ with  metric $\eta_{\mu\nu}=\textrm{Diag}(d-1,-1,\ldots,-1)$.  Thus the
appearance of a Minkowski product is to be expected.

As a consequence of Lemma \ref{unitary}, unitary transforms, since they
leave $\textrm{Herm}^+_d(\mathbb{C})$ invariant, yield rotations
which leave $C$ (globally) invariant. This fact deserves to be
analysed in detail to understand the geometry of $C$.  As Unitary transforms act transitively on
pure states, $S^+$ is the closed convex hull
of a homogeneous subspace $\phi(\mathcal{P})$ of $\partial S^{d^2-2}$. For the moment however, we shall
consider the geometric representation of general quantum operations in $C$.

\subsection{Generalized measurements}
\label{generalized measurements}
We call a generalized measurement \cite{Nielsen} a finite set $\{M_m\}_m$ of complex $d\times
d$ matrices which satisfy :
$\sum_{m} M_m^{\dagger}M_m=\mathbb{I}$. The set of
$\{E_m\}_m=\{M_m^{\dagger}M_m\}_m$ defines a Positive Operator Valued
Measure (POVM), as $E_m\in\textrm{Herm}^+_d(\mathbb{C})$ and
$\sum_{m}E_m=\mathbb{I}$. Given a quantum state or density matrix
$\rho\in D$, the generalized 
measurement $\{M_m\}_m$ on $\rho$ yields outcome $m$ with probability
$p(m)=\trace(E_m\rho)$, and if outcome $m$ occurs, the post
measurement state is
$\rho'_m=(1/ \trace(E_m \rho))(M_m \rho M_m^{\dagger})$. We shall call $\rho_m=
M_m \rho M_m^{\dagger}\in \textrm{Herm}^+_d(\mathbb{C})$ the \emph{unrescaled} post-measurement state. \\
Recall that any complex matrix can be polar-decomposed into a product of a
unitary matrix and a positive matrix. For all $m$, there exists
$U_m\in U(d)$ and $A_m\in \textrm{Herm}^+_d(\mathbb{C})$ such
that $M_m=U_mA_m$. As $E_m=M_m^{\dagger}M_m=A_mA_m$, $A_m=\sqrt{E_m}$,
the positive square root of $E_m$. Using this polar decomposition,
$\rho'_m$ is represented in the cone $C$ by
\begin{align*}
\ul{ \rho'_m} &\equiv \phi(\rho'_m)= \frac{1}{\trace(E_m\rho)}\phi(U_m
\sqrt{E_m}\rho\sqrt{E_m} U_m^{\dagger}) \\
&= \frac{1}{\trace(\sqrt{E_m}\rho\sqrt{E_m})}\psi(U_m)(\ul{\sqrt{E_m}\rho\sqrt{E_m}})
\end{align*}
Thus when outcome $m$ occurs, the post-measurement state $\ul{ \rho'_m}$ of $\{M_m\}_m$ is the same as
that of $\{\sqrt{E_m}\}_m$ up to a rotation $\psi(U_m)$, and similarly
for the unrescaled states. As a consequence we shall consider the geometrical
effects of generalized
measurements $\{\sqrt{E_m}\}_m$ where $E_m$ and $\sqrt{E_m}$ are in
$\textrm{Herm}^+_d(\mathbb{C})$ and $\sum_{m}E_m=\mathbb{I}$, bearing
in mind that the most general measurements just involve rotations on the
post-measurement state vectors. For example, in section \ref{part three}, Eve is free to perform
unitary transforms on her post-measurement states, and  can decide
this according to the outcome $m$. The procedure we use to find the
Disturbance is to first measure with $\{\sqrt{E_m}\}_m$ and then
maximise on Unitary
transforms acting upon post-measurement states. Using
the conal representation, both sets of vectors $\{\ul{E_m}\}_m$ and
$\{\ul{\sqrt{E_m}}\}_m$ are in $C$, and
$\sum_{m}\ul{E_m}=(d,0,\ldots,0)$. This enables us to represent
elements of a measurement inside $C$, and visualize the action of a
particular non-trace-preserving operation 
$\sqrt{E_m}$ on a given density matrix $\rho$, in other words find
$\ul{\rho_m}$ in terms of $\ul{E_m}$ or $\ul{\sqrt{E_m}}$.

\subsection{Quantum operations represented in $C$}

One might wonder here why not just rescale all the post-measurement
states and only consider the density matrices $\rho'_m$. The reason
for \emph{not} doing so is that the unrescaled states encode extra
information: their ``height'' in
the cone, the first component $\ul{\rho_{m}}_0=\trace(E_m\rho)$, is
simply the probability of their outcomes. Under a given generalized
measurement, post-measurement vectors with identical first components
are equiprobable. Thus the  sections of $C$ of constant $\lambda_0$ have a
clear physical interpretation. We shall need the
following simple properties:
\begin{Lem}
\label{trace positivity}
For $A\in \textrm{Herm}_d(\mathbb{C})$ and
$B,C\in\textrm{Herm}^+_d(\mathbb{C})$, 
\begin{align*}
\trace(BC)&\geq 0  \\ 
\trace(BABA)&\geq 0
\end{align*}
\end{Lem}
\textbf{Proof:} Let $B=\sqrt{B}\sqrt{B}$, then
$\trace(BC)=\trace(\sqrt{B}C\sqrt{B})\geq 0$ since
$\sqrt{B}C\sqrt{B}\in \textrm{Herm}^+_d(\mathbb{C})$. Then polar
decompose $A$ into $A=U\left| A\right|$, with $U$ unitary and
$\left |A\right| \in\textrm{Herm}^+_d(\mathbb{C})$. As
$A\in \textrm{Herm}_d(\mathbb{C})$, $A=\left| A\right| U^{\dagger}=A^{\dagger}$,
and 
\begin{displaymath}
\trace(BABA)=\trace(BU\left|A\right|B\left| A\right|
U^{\dagger})=\trace(U^{\dagger}BU\left| A\right| B\left|A \right|)
\end{displaymath}
This is non-negative by the previous result since $U^{\dagger}BU,
\left| A\right| B\left| A\right| \in
\textrm{Herm}^+_d(\mathbb{C})$. $\quad \Box$

Unitary transforms
induce rotations in $C$, and generalized measurements have the
following geometric properties:
\begin{Prop}
\label{symmetricity}
The linear transforms
$\psi(\sqrt{E_m}):\ul \rho\mapsto \ul{\rho_m}$  associated to a
generalized measurement $\{\sqrt{E_m}\}_m$ correspond to  
real symmetric matrices which are positive.  They individually map
$\mathcal{C}$ into itself. In addition, for any generalized pure state $\theta$,
$\psi(\theta)$ maps C into $\mathcal{C}$.\\
The probability of outcome $m$ for a quantum system in state $\rho$ is given by
\begin{equation}
\label{probability}
p(m)=\frac{1}{d}\ul{E_m}.\ul{\rho}
\end{equation}
\end{Prop}
$\mathbf{Proof:}$ By using (\ref{components}) successively, we have
\begin{align}
\ul{\rho_m}_{\mu}&=\trace(\sqrt{E_m}\rho\sqrt{E_m}\tau_{\mu}) \nonumber
\\
\label{matrix coefficients} &=\frac{1}{d}\trace(\sqrt{E_m}\tau_{\nu}\sqrt{E_m}\tau_{\mu})\ul{\rho}_{\nu} \\
&\equiv M^m_{\mu\nu}\,\ul \rho_{\nu} \nonumber
\end{align}
Clearly $ M^m_{\mu\nu}$ is  real symmetric by cyclicity of the trace
and the fact that 
$\sqrt{E_m}\tau_{\nu}\sqrt{E_m}$ and $\tau_{\mu}$ are hermitian. (Actually
 $\psi(A)$ is real for any complex $d\times d$ matrix A, and real
symmetric for any A hermitian). Let $\ul
v=(\ul v_{\mu})\in \mathbb{R}^{d^2}$. Using (\ref{matrix
coefficients}) we get 
\begin{align*}
\ul v^T\psi(\sqrt{E_m})\ul v&= \ul v_{\mu}M^m_{\mu\nu}\ul v_{\nu}
=\frac{1}{d}\ul
v_{\mu}\trace(\sqrt{E_m}\tau_{\nu}\sqrt{E_m}\tau_{\mu})\ul v_{\nu}\\
&=\frac{1}{d}\trace(\sqrt{E_m}(\ul v_{\nu}\tau_{\nu})\sqrt{E_m}(\ul
v_{\mu}\tau_{\mu})) \geq 0
\end{align*}
This follows from Lemma \ref{trace positivity} since $\ul
v_{\mu}\tau_{\mu}\in  \textrm{Herm}_d(\mathbb{C})$. Hence
$M^m_{\mu\nu}$ is a positive real (symmetric) matrix.\\
The properties on purity simply follow from general facts on quantum
operations on density matrices which remain true for
generalized density matrices:\\
For $|u><u|$ and $|v><v|$ generalized pure states, for
any $A$ complex $d\times d$ matrix and any generalized density
matrix $\rho$,  
\begin{align}
A|u><u|A^{\dagger}& =|A\,u><A\,u| \quad \textrm{and} \nonumber \\
\label{pure measurement}
|v><v|\rho|v><v|&=<v|\rho|v>|v><v|
\end{align}
are generalized pure states. Relation (\ref{probability}) follows from (\ref{components})
and $\trace(\sqrt{E_m}\rho\sqrt{E_m})=\trace(E_m\rho)$. $\quad \Box$\\   

The following properties will help to give a geometrical intuition of
the action of the $\psi(\sqrt{E_m})$'s. For $\sqrt{E_m}=|v><v|$ pure,
$<v|\rho|v>=\trace(\sqrt{E_m}\rho)=(1/d)\ul{\sqrt{E_m}}.\ul\rho$. Thus
using (\ref{pure measurement}):
\begin{displaymath}
\psi(\sqrt{E_m})\ul\rho=\frac{1}{d}(\ul{\sqrt{E_m}}.\ul\rho)\ul{\sqrt{E_m}}
\end{displaymath}
So $\psi(\sqrt{E_m})$ is as was expected a non-normalized projection. For
any $\sqrt{E_m}\in \textrm{Herm}^+_d(\mathbb{C})$ , 
the $d^2$ eigen-vectors $\ul{v^{\sigma}}$ of $\psi(\sqrt{E_m})$ with
eigen-values $\lambda^{\sigma}$ correspond to
$d^2$ hermitian matrices $M^{\sigma}\equiv \phi^{-1}(\ul{v^{\sigma}})$ which satisfy
$\sqrt{E_m}M^{\sigma}\sqrt{E_m}=\lambda^{\sigma}M^{\sigma}$ (no
summation). As a consequence, if $\ul{\rho}\in C$ is such an eigenvector, then the rescaled density
matrix $\rho$ is such that $\rho=\rho'_m$, i.e. $\rho$ is unchanged if outcome $m$
occurs.\\ 

We now give the general expressions for $\rho_m$ in terms of
$\rho\equiv(1/d) \rho_{\mu}\tau_{\mu}$ and
$\sqrt{E_m}\equiv(1/d)\E_{\nu}\tau_{\nu}$, where we drop the
index $m$ and do not underline the components of the vectors
$\ul{\rho}$ and $\ul{\sqrt{E_m}}$ for
convenience. By definition:
\begin{displaymath}
\rho_m=\frac{1}{d^3}\E_{\mu}\rho_{\nu}\E_{\sigma}\tau_{\mu}\tau_{\nu}\tau_{\sigma}
\end{displaymath}
Expanding this using $\tau_0=\mathbb{I}$ and grouping the products of
the $\tau_i$'s in hermitian terms, we easily derive:
\begin{align}
\rho_m&=\frac{1}{d^3}\Big\{
\E_0\rho_0\E_0\mathbb{I}+(2\E_0\rho_0\E_i+\E_0\rho_i\E_0)\tau_i
\nonumber  \\
&\;
+(\frac{1}{2}\E_i\rho_0\E_j+\E_0\rho_i\E_j)(\tau_i\tau_j+\tau_j\tau_i)
\nonumber \\
\label{ttt}&\;
+\frac{1}{2}\E_i\rho_j\E_k(\tau_i\tau_j\tau_k+\tau_k\tau_j\tau_i)\Big\}
\end{align}
To push the general $d$-dimensional analysis further, we need a particular choice of
$\tau_i$'s whose anti-commutation relations are convenient. This is
subject to current work. We now treat in full detail the $d=2$ (one
qubit) case and apply our geometric approach to a challenging quantum
information theoretical problem.

\section{The Qubit Case Pushed Further}
\label{part two}
Applied to qubit states the representation yields two of the most
familiar objects in fundamental physics: the $2\times 2$ density
matrices yield a Minkowskian future-light-cone in
$\mathbb{E}^{1,3}$ whose vertical sections are nothing but Bloch
spheres. The correspondence between light-like vectors and fully
determined spins is puzzling, but it requires a little more than
one qubit to be investigated further. Meanwhile in this simple
case we are able to give explicit coordinates for states posterior
to non trace-preserving quantum operations. These formulae remain
simple provided Minkowskian products are introduced alongside the
Euclideans. They constitute a sufficient armoury to deal, using
only four-vectors, with the most general evolutions to happen on a
qubit.
\subsection{The Cone and the Bloch Sphere}
A suitable Hilbert-Schmidt orthogonal basis for $2\times 2$
traceless hermitian matrices is given by the set of Pauli
matrices:
\begin{align*}
\tau_1&=\mathbf{X}=\left(\begin{array}{cc}
 0 &1 \\
 1 &0\end{array}\right)\\
\tau_2&=\mathbf{Y}=\left(\begin{array}{cc}
 0 &-i \\
 i &1\end{array}\right)\\
\tau_3&=\mathbf{Z}=\left(\begin{array}{cc}
 1 &0 \\
 0 &-1\end{array}\right)
\end{align*}
Together with the identity
\begin{align*}
\tau_0&=\mathbb{I}= \left(\begin{array}{cc}
 1 &0 \\
 0 &1\end{array}\right)
\end{align*}
one may express any $2\times 2$ hermitian matrix as a sum
$A=\frac{1}{2}\ul A_{\mu} \tau_{\mu}$ with the $\ul A_{\mu}$'s real. The
positivity conditions for those matrices turns out simple.
\begin{Lem}
The cone of positive hermitian matrices
$\textrm{Herm}_2^+(\mathbb{C})$ is isomorphic to the following
cone of revolution in $\mathbb{R}^4$:
\begin{align*}
\Gamma&=\{(\lambda_\mu) \in \mathbb{R}^4 \,/\, \lambda_0^2
-\sum_{i=1}^3 \lambda_i^2 \geq 0, \lambda_0 \geq 0\}
\end{align*}
Generalized pure states lie on the boundary of $\Gamma$.
\end{Lem}
\textbf{Proof}: The eigenvalues of $A$ are given by
$\lambda_\pm=\frac{1}{2}(\ul A_0 \pm \sqrt{\ul A_i \ul A_i})$.
$A$ is positive if and only if $\lambda_+ \lambda_- \geq 0$
and $\lambda_+ + \lambda_- \geq 0$. This is equivalent to: 
\begin{align}
\eta_{\mu\nu}\ul A_{\mu} \ul A_{\nu} \geq 0 \, \wedge \, \ul A_0 \geq 0
\label{positivity condition}
\end{align}
with $\eta_{\mu \nu}=\textrm{Diag}(1,-1,-1,-1)$.
The purity condition is an obvious consequence of Proposition
\ref{cone}. {$\quad \Box$}

Thus the generalized (not necessarily normalized) density matrices
of a qubit cover the whole Minkowskian future-light-cone in
$\mathbb{E}^{1,3}$. Taking a vertical cross-section of the cone
is equivalent to fixing the trace $\ul A_0$ of the density matrix, which might be thought of physically as the \emph{overall
probability of occurrence} for the state. By doing so we are left
with only the spin degrees of freedom along $\mathbf{X}$, $\mathbf{Y}$, $\mathbf{Z}$,
therefore each vertical cross-section is a Bloch sphere with
radius $a=\ul A_{0}$.


The ability to represent states with different traces is
convenient when dealing with quantum ensembles
$\{(p_x,\rho_x)\}_x$. When we seek to represent non
trace-preserving quantum operations the feature becomes absolutely
crucial.
\subsection{The Post-measurement State}
As we have seen in subsection \ref{generalized measurements}, the most general quantum operation can be
described as $\{M_m\}_m=\{U_m \sqrt{E_m}\}_m$ with $U_m$ unitary and
$\sqrt{E_m}$ positive (the only exta feature Kraus operators allow is
the possibility to ignore one's knwoledge of some measurement
outcomes, but in our setting this is easily catered for by adding up
the undistinguished non-normalized post-measurement states). While the action of $U_m$ is
well understood in terms of four-vectors (as a mere rotation in
the Bloch Sphere, see Lemma \ref{unitary}), the authors of this paper are not
aware of a solid geometrical framework for representing the
effects of $\sqrt{E_m}$ - other than the one presented here. In Lemma \ref{arhoa}, if $A\equiv\sqrt{E_m}$ while $\rho$ is
the initial state, then $A\rho A$ stands for the (not
renormalized) `post-measurement' state when outcome $m$ has
occurred (up to a unitary evolution $U_m$).
\begin{Lem} \label{arhoa} Let $A$ and $\rho$ be two matrices in
$\textrm{Herm}_2^+(\mathbb{C})$. Then:
\begin{align}
 A\rho A =&\frac{1}{8}[-\ul \rho_0 (\eta_{\mu \mu'} \ul
A_{\mu}
\ul A_{\mu'}) + 2 \ul A_0(\ul A.\ul \rho)] \tau_0 \nonumber\\
&+\frac{1}{8}[\ul \rho_1 (\eta_{\mu \mu'} \ul A_{\mu}
\ul A_{\mu'}) + 2 \ul A_1 (\ul A.\ul \rho)] \tau_1 \nonumber\\
&+\frac{1}{8}[\ul \rho_2 (\eta_{\mu \mu'} \ul A_{\mu}
\ul A_{\mu'}) + 2 \ul A_2 (\ul A.\ul \rho)] \tau_2 \nonumber\\
&+\frac{1}{8}[\ul \rho_3 (\eta_{\mu \mu'} \ul A_{\mu} 
\ul A_{\mu'}) + 2 \ul A_3 (\ul A.\ul \rho)] \tau_3 \nonumber \\
=& \frac{1}{8}[\eta_{\nu \nu'}\ul \rho_{\nu} ( \eta_{\mu \mu'}
\ul A_{\mu} \ul A_{\mu'}) + 2 \ul A_{\nu'} ( \ul A.\ul \rho)]
\tau_{\nu'} \label{arhoaform}
\end{align}
\end{Lem}
\textbf{Proof}: Consider
\begin{align*}
\ul A& =   [\begin{array}{cccc} \alpha
&\beta &\gamma &\delta \end{array}]\\
\ul \rho& =   [\begin{array}{cccc} a &x &y &z
\end{array}]
\end{align*}
We have:
\begin{align}
A\rho A =&\frac{1}{8}[a(\alpha^2+\beta^2+\gamma^2+\delta^2)+2\alpha(\beta x+\gamma
y+\delta z)] \tau_0 \nonumber\\
&+\frac{1}{8}[x(\alpha^2+\beta^2-\gamma^2-\delta^2)+2\beta(\alpha a+\gamma
y+\delta z)] \tau_1 \nonumber\\
&+\frac{1}{8}[y(\alpha^2-\beta^2+\gamma^2-\delta^2)+2\gamma(\alpha a+\beta x+\delta z)] \tau_2 \nonumber\\
\label{arhoabadform}&+\frac{1}{8}[z(\alpha^2-\beta^2-\gamma^2+\delta^2)+2\delta(\alpha a+\beta
x+\gamma y)] \tau_3
\end{align}
This formula can be be obtained either by brute force calculation
using the Pauli multiplication relations, or by exploiting the
fact that Pauli matrices form a Clifford Algebra I.e. $\{\tau_i
,\tau_j\}=2\delta_{ij}\tau_0$ together with equation (\ref{ttt}). Regrouping the
terms gives formula (\ref{arhoaform}). $\quad \Box$
\begin{Cor}
Let $A$ and $\rho$ be two matrices in
$\textrm{Herm}_2^+(\mathbb{C})$. $A\rho A$ can be expressed as a
linear combination of $\rho$, $A$ and the Identity:
\begin{align*}
A\rho A &= \frac{1}{2^4} (\ul A.\ul \rho) \, A + \frac{1}{2^5}
(\eta_{\mu \mu'} \ul A_{\mu} \ul A_{\mu'}) (\rho - \rho_0 \tau_0)
\nonumber
\end{align*}
\end{Cor}
This last corollary provides much geometrical insight on non
trace-preserving quantum operations. We find that the effect of
$\sqrt{E_m}$ is not that difficult to visualize: the resulting
state is a weighted sum of $\sqrt{E_m}$, the initial state and
the identity, with real coefficients.

It is a somewhat strange fact that the structure equation (\ref{arhoabadform}) does
not become apparent until one brings the Minkowskian product to
the rescue. The spurious appearance of special relativistic
products in quantum mechanics bears some explanation in this setting however,
since the Minkowski metric is intrinsically related to the
characteristic function of pointed cones of revolution.

Finally it is important to notice that the results expressed in
these two last subsections are invariant under any orthogonal change of basis
$\{\tau_i\}_i$. This is because rotations about the vertical axis
leave the Minkowskian product invariant. The Pauli matrices have been
helpful in computing those results, but from now and in the rest
of the paper we may consider ourselves in the more general setting
of section \ref{part one}.
\subsection{Square and Square Root}
In our quest towards representing non trace-preserving quantum
operations in the cone we have managed to obtain the probability
of occurrence $p(m)$ in terms of $\ul E_m$ (Proposition \ref{symmetricity}). In the
previous subsection we have also worked out the evolved state $\ul{\rho_m}$, but unfortunately this was done in terms of
$\ul{\sqrt{E_m}}$. In order to deal fully with these operations in
the Cone formalism we need to understand ways of switching back
and forth from $\ul E_m$ to $\ul{\sqrt{E_m}}$. The next Lemma is a
direct consequence of equation (\ref{arhoaform})  when $\rho=\mathbb{I}$.
\begin{Lem} The square of a matrix $A$ in $\textrm{Herm}_2^+(\mathbb{C})$ is given by:
\begin{align*}
A^2 &=\ul A_0 \, A - \frac{1}{4}(\eta_{\mu \nu}\ul A_{\mu}\ul
A_{\nu}) \tau_0
\end{align*}
Inversely the square root operation obeys:
\begin{align*}
\sqrt{A} &=\frac{1}{r}(A + \frac{1}{2}\sqrt{\eta_{\mu \nu}\ul
A_{\mu}\ul A_{\nu}} \,
\tau_0) \\
\mbox{with:} \quad r &=\sqrt{\ul A_0+\sqrt{\eta_{\mu \nu}\ul A_{\mu}\ul
A_{\nu}}}\nonumber\\
\end{align*}
\end{Lem}
Note that $A$ is proportional to $\sqrt{A}$ if and only if $A$ is
generalized pure or $A\propto{\mathbb{I}}$.\\
But when we seek to express a function of $\ul E_m$ in terms of
$\ul{\sqrt{E_m}}$ (or the reverse) the next formulae become
convenient.
\begin{Lem} Let $A$ and $\rho$ be two matrices in
$\textrm{Herm}_2^+(\mathbb{C})$. The following relations hold:
\begin{align*}
\eta_{\mu \nu}\ul{\sqrt{A}}_{\mu}\ul{\sqrt{A}}_{\nu}&= 2\sqrt{\eta_{\mu \nu}\ul A_{\mu} \ul A_{\nu}} \\
\ul A^2 .\ul \rho &= \ul A_0 (\ul A.\ul \rho) - \frac{1}{2}\ul \rho_0 (\eta_{\mu \nu}\ul A_{\mu} \ul A_{\nu}) \\
\ul{\sqrt{A}}.\ul \rho &= \frac{1}{r}(A.\ul \rho +\ul \rho_0 \sqrt{\eta_{\mu \nu}\ul A_{\mu} \ul A_{\nu}}) \\
\mbox{with:}\quad r &=\sqrt{\ul A_0+\sqrt{\eta_{\mu \nu}\ul A_{\mu}\ul
A_{\nu}}}\nonumber\\
\end{align*}
\end{Lem}
On the whole taking the square root of $\ul E_m$ is not so easy.
It would be much more convenient if we could make all calculations
in terms of $\ul E_m$, with the added advantage condition:  
\begin{align}
\sum_m E_m &= 2 \tau_0 \label{sum to identity}
\end{align} 
is easily visualized.
Results in the
following subsection are most useful for this purpose.
\subsection{Inner Products Through Quantum Operations}
Consider two states $\rho^0$, $\rho^1$. Suppose they undergo a
quantum operation $\{M_m\}_m=\{U_m \sqrt{E_m}\}_m$ and outcome $m$
occurs. Rather than seeking the coordinates of the rescaled
post-measurement states ${\rho^0_m}'$ and ${\rho^1_m}'$, we may be
interested in their positions \emph{relative to one another}. Note
this subsection reuses a number of notational conveniences
introduced in section \ref{part one}.
\begin{Lem} Let $\rho^0$,$\rho^1$ be two initial states in
$\textrm{Herm}_2^+(\mathbb{C})$ and $\sqrt{E_m}$ a measurement element
in $\textrm{Herm}_2^+(\mathbb{C})$. The inner products of the
post-measurement states satisfy:
\begin{align}
\label{four scalar product}
&\ul{\rho^0_m} .\ul{\rho^1_m} = \\
&\frac{1}{4}[2(\ul{E_m}.\ul{\rho^0})(\ul{E_m}.\ul{\rho^1})-(\eta_{\mu
\mu'} \ul{E_m}_{\mu}\ul{E_m}_{\mu'})(\eta_{\nu
\nu'}\ul{\rho^0}_{\nu}\ul{\rho^1}_{\nu'})]\nonumber\\
&\ul{{\rho^0_m} '}.\ul{{\rho^1_m} '}=2-\frac{(\eta_{\mu \mu'}
\ul{E_m}_{\mu}\ul{E_m}_{\mu'})(\eta_{\nu
\nu'}\ul{\rho^0}_{\nu}\ul{\rho^1}_{\nu'})}{(\ul{E_m}.\ul{\rho^0})(\ul{E_m}.\ul{\rho^1})}\nonumber\\
&\overrightarrow{\rho^0_m}.\overrightarrow{\rho^1_m} = \nonumber \\
&\frac{1}{4}[(\ul{E_m}.\ul{\rho^0})(\ul{E_m}.\ul{\rho^1})-(\eta_{\mu
\mu'} \ul{E_m}_{\mu}\ul{E_m}_{\mu'})(\eta_{\nu
\nu'}\ul{\rho^0}_{\nu}\ul{\rho^1}_{\nu'})]\nonumber\\
&\overrightarrow{{\rho^0_m}'}.\overrightarrow{{\rho^1_m}'}=1-\frac{(\eta_{\mu
\mu'} \ul{E_m}_{\mu}\ul{E_m}_{\mu'})(\eta_{\nu
\nu'}\ul{\rho^0}_{\nu}\ul{\rho^1}_{\nu'})}{(\ul{E_m}.\ul{\rho^0})(\ul{E_m}.\ul{\rho^1})}
\label{three scalar product}
\end{align}
\end{Lem}
\textbf{Proof:}
By using (\ref{isometry}) we have:
\begin{align*}
\ul{\rho^0_m} .\ul{\rho^1_m} &=\ul{\sqrt{E_m} \rho^0 \sqrt{E_m}}.\ul{\sqrt{E_m} \rho^1
\sqrt{E_m}}\\
&=2 Tr(\sqrt{E_m} \rho^0 \sqrt{E_m}\sqrt{E_m} \rho^1
\sqrt{E_m})\\
&=2 Tr(E_m \rho^0 E_m\rho^1)\\
&= \ul{E_m \rho^0 E_m}.\ul{\rho^1}
\end{align*}
From there we readily obtain equation (\ref{four scalar product}) by
applying equation (\ref{arhoaform}) once. $\quad \Box$

By letting $\rho^0=\rho^1=\rho$ in the above Lemma we get:
\begin{align}
||\ul{\rho_m}||^2&=\frac{1}{4}[2(\ul{E_m}.\ul{\rho})^2-(\eta_{\mu
\mu'} \ul{E_m}_{\mu}\ul{E_m}_{\mu'})(\eta_{\nu
\nu'}\ul{\rho}_{\nu}\ul{\rho}_{\nu'})]\nonumber\\
||\ul{{\rho_m}'}||^2&=2-\frac{(\eta_{\mu \mu'}
\ul{E_m}_{\mu}\ul{E_m}_{\mu'})(\eta_{\nu
\nu'}\ul{\rho}_{\nu}\ul{\rho}_{\nu'})}{(\ul{E_m}.\ul{\rho})^2}\nonumber\\
||\overrightarrow{{\rho_m}}||^2&=\frac{1}{4}[(\ul{E_m}.\ul{\rho})^2-(\eta_{\mu
\mu'} \ul{E_m}_{\mu}\ul{E_m}_{\mu'})(\eta_{\nu
\nu'}\ul{\rho}_{\nu}\ul{\rho}_{\nu'})]\nonumber\\
||\overrightarrow{{\rho_m}'}||^2&=1-\frac{(\eta_{\mu \mu'}
\ul{E_m}_{\mu}\ul{E_m}_{\mu'})(\eta_{\nu
\nu'}\ul{\rho}_{\nu}\ul{\rho}_{\nu'})}{(\ul{E_m}.\ul{\rho})^2}
\label{three ned norm}
\end{align}
Equation (\ref{three ned norm}) clearly exhibits the general property we stated in
Proposition \ref{symmetricity}: that is if the initial state is
generalized pure ($\eta_{\mu \mu'} \ul \rho_{\mu} \ul
\rho_{\mu'}=0$) or the measurement is generalized pure ( $\eta_{\mu
\mu'} \ul{E_m}_{\mu} \ul{E_m}_{\mu'}=0$) then we have $||\overrightarrow{{\rho_m}'}||=1$
(pure), which implies that $\rho_m$ is generalized
pure.

The above lemma enables us to determine all the \emph{relative
positions} (angles and norms) of quantum states using relatively compact formulae
which do not involve $\sqrt{E_m}$. It is only when the coordinates
of each post-measurement state are required that one needs to take
the impractical square root of $E_m$. But remember we are allowed an arbitrary
rotation $U_m$ in order to complete the quantum operation. This means we
have full freedom to fix the absolute coordinates at will (so long
as the relative positions are respected). 

Most Quantum Information Theoretical problems seek to evaluate the limits of
quantum operations, e.g. quantum cloning \cite{Cerf}, distinguishability \cite{distinguishability},
Infomation Gain versus Disturbance tradeoff \cite{Banaszek}. In these situations the
precise individual coordinates of the states
after $Ad_{\sqrt{E_m}}$ tend not to matter; usually they will need to be rotated
anyhow into a position which optimizes the fidelity measure in
question. What counts is the relative position of the post-measurement
states. Therefore these problems can be treated comfortably in our
framework. Section \ref{part three} provides a good example of such an application. 

There are, however, some rare situations where we would like to see
quantum operations act step by step, yielding precise coordinates - 
instead of just fixing the coordinates of the final state as we would
do in order to avoid taking the square root of ${E_m}$. This is
the case for instance in quantum complexity, where one needs an
appreciation of how many basic computational operations it takes to
accomplish some calculation. Yet in this type of problems it turns out that the basic
operations can be taken to be unitary operators, with measurements
only performed at the end (principle of delayed measurement \cite{Nielsen}). Therefore these
scenarios may still be analyzed comfortably within our conal
representation: the basic unitary operators will just be a set of chosen real
orthogonal rotations, and the final measurement statistics will be
evaluated straight from $E_m$.

\section{Application: Information gain versus Disturbance Tradeoff}
\label{part three}
The following idealized scenario captures a key situation for any
quantum cryptographic protocol:

Alice owns a random variable $X= \{(\frac{1}{2},0),(\frac{1}{2},
1)\}$. According to the outcome $x$ she prepares either
$|{\psi}_0\rangle$ or $|{\psi}_1\rangle$, i.e. she runs $|x\rangle
|0\rangle \stackrel{U}{\rightarrow} |x\rangle |{\psi}_x\rangle $.
Eve knows $U$ and the distribution $X$, but not the particular
outcome Alice has drawn. Later Eve gains access to $|{\psi }_x
\rangle$ and may use of this opportunity to try and learn about
$x$. How much she learns is quantified using Information
theoretical notions. Even though Alice has had to expose $|{\psi
}_x \rangle$, still she really wanted to keep $x$ secret. But now
she gets a chance of checking upon Eve's honesty - by asking her
to return $|{\psi }_x \rangle$. Suppose Eve's measurement and
further manipulations have modified $|{\psi}_x\rangle$ into ${\rho
}_x$. Alice then measures $\{{\ket{\psi_x}\bra{\psi_x},
Id-\ket{\psi_x}\bra{\psi_x}}\}$ and has a probability $1-\langle
{\psi}_x | {\rho}_x | {\psi }_x \rangle$ of detecting the felony.

The point is that most quantum cryptographic protocols rely upon the fact that
Eve cannot eavesdrop a state without causing it an irreversible,
detectable damage.  In spite of their central role, Information
Gain versus Disturbance tradeoffs upon discrete ensemble states
remain largely unknown, due to the mathematical difficulties they
raise. In 1995 Fuchs and Peres accomplished the mathematical feat
of obtaining an analytic formula for the above case of two
non-orthogonal states. But the method they used relies upon a
number of ``plausible'' assumptions - and does not provide a
geometrical intuition of what the family of optimal measurements
looks like.

The Cone, by enabling a \emph{per outcome} geometrical
representation of generalized measurements,  permits us to
overcome some of these shortcomings and greatly facilitate the
derivation of Fuchs and Peres' formula. We hope this illustrates the power
of the geometrical framework developed in this paper.

\subsection{Information Contribution, Disturbance Contribution}
Suppose the $\{|\psi_x\rangle\}_{x=0,1}$ states Alice prepares verify the
following basic relations:
\begin{align*}
\underline{v^x} & =  \phi(|\psi_x\rangle\langle\psi_x|) \\
\underline{v^0}.\underline{v^1} & =  d = \sqrt{1-c^2} \\
\end{align*}
By choosing a suitable basis in the Bloch Sphere and since the
$\{|\psi_x\rangle\}_x$ are pure we may fix:
\begin{align}
\underline{v^0} & =  [\begin{array}{cccc} 1 \;&c \; \, \: \! &d
&0\end{array}] \label{v0 coord}\\
\underline{v^1} & =  [\begin{array}{cccc} 1 &-c &d &0\end{array}]
\label{v1 coord}
\end{align}

The most general thing Eve can ever do is to attack the states
with a measurement $\{M_m\}_m$. This procedure is equivalent to
first measuring $\{\sqrt{E_m}\}_m$, and then, conditional to $m$,
applying the unitary transformation
$U_m$, with $E_m$ and $U_m$ defined as in subsection \ref{generalized measurements}. \\
It is rather interesting to observe that the second step has no
other use but to ``repair'' the post-measurement states as much as
is possible. The first step on the other hand may partially
destroy the initial states so as to collect the Information
Eve seeks. This is the step we now study in order to quantify her
Information Gain.

Let $Y$ be the random variable arising from the measurement
outcomes, i.e. $Y=\{(p(m),m)\}_m$. Eve's Information Gain is given
by:
\begin{align*}
I & =  H(X:Y) = H(Y)-H(Y|X)   \nonumber\\
& =  \sum_m p(m)\log(p(m))-\sum_{x,m}p(x,m)\log(p(m|x)) \nonumber\\
& \equiv  \sum_m I_m \quad \mbox{with} \\
I_m& =  p(m)\log(p(m))-\sum_{x}p(x,m)\log(p(m|x))
\end{align*}
$I_m$ must be understood as the \emph{Information Contribution}
brought by the measurement element:
\begin{align*}
\epm & =  [\begin{array}{cccc} \alpha &\beta &\gamma
&\delta\end{array}]=\phi(E_m)
\end{align*}
By making use of the relations (\ref{v0 coord}),(\ref{v1 coord}) and (\ref{probability}) one can express $I_m$
geometrically in terms of scalar products in the cone:
\begin{align}
I_m =& -(p_m+q_m)\log(p_m+q_m)\nonumber\\
&+p_m \log(2p_m)+q_m\log(2q_m)\nonumber\\
\mbox{with}\quad p_m = &\frac{\alpha+\beta c+\gamma
d}{4}=\frac{\epm.\vo}{4}\equiv p(0,m) \label{pm} \\
q_m= &\frac{\alpha-\beta c+\gamma
d}{4}=\frac{\epm.\vi}{4}\equiv p(1,m) \label{qm}
\end{align}
Notice that if $\epm$ is orthogonal to $\vi$ (resp. $\vo$) then
$I_m=p_m$ (resp. $q_m$). Such a measurement element may be said to be ``all or
nothing'': it brings a whole bit of information when it occurs, but
does so only with probability $p_m$ (resp. $q_m$). Taken individually
these measurement elements seem ideal: they fully identify
$|\psi_x\rangle$ and thus they let you reconstruct the initial state
perfectly, with no disturbance at all. The downside is that failure to
occur comes at a high price. In order to verify the condition (\ref{sum to
identity}) the other measurement elements generally become
rather inefficient with respect to the tradeoff. The family of the optimizing
$\{M_m\}_m$ is not constructed in such simple ways.\\

Next we seek an expression of the \emph{Disturbance Contribution}
brought by each measurement element. For this purpose we must
first assume outcome $m$ has occurred. Eve knows it, and now she
will try to maximize her chances of fooling Alice by applying a
carefully tailored unitary evolution $U_m$. First we will give $D_m$
as a function of $U_m$, and next proceed to the maximization which
determines $U_m$. Remember that upper indices $x$ distinguish initial
states, while lower indices $m$ specify the measurement outcome. 
\begin{align*}
p(\textrm{fool}|m) & =\sum_x p(x|m) \trace(\ket{\psi_x}\bra{\psi_x}U_m{{\rho^x_m}'} U_m)\\
 & \equiv  \frac{\sum_x p(x|m)\vx.\ul{ r^x_m}}{2} \\
 & =  \frac{1+ \sum_x p(x|m)\overrightarrow{v^x}.\overrightarrow{r^x_m}}{2}\quad \mbox{where} \nonumber \\
{\ul r^x_m} & =  [\begin{array}{cc} 1 &{\overrightarrow{r^x_m}}\end{array}]\equiv \phi(U_m{{\rho^x_m}'} U_m)\\
& =  \frac{\phi (U_m \sqrt{E_m} \ket{\psi_x}\bra{\psi_x}
\sqrt{E_m} U_m)}{p(m|x)}
\end{align*}
Negating back to the Disturbance we obtain:
\begin{align*}
D & =  \sum_m D_m \quad \mbox{with} \\
D_m & =  p(\neg \textrm{fool},m) \nonumber \\
& =  \frac{p(m)- \sum_x p(x,m)\overrightarrow{v^x}.\overrightarrow{r^x_m}}{2}\\
& =  \frac{p(m)- \sum_x p(x,m)\|\overrightarrow{v^x}\|\,\|\overrightarrow{r^x_m}\|
\cos\widehat{(\overrightarrow{v^x},\overrightarrow{r^x_m})}}{2} \nonumber
\end{align*}
In our scenario the $\{\ket{\psi_x}\}_x$ are pure. Thus by Lemma
\ref{symmetricity} or equation (\ref{three ned norm}) we have $\|\overrightarrow{v^x}\|\,\|\overrightarrow{r^x_m}\|=1$. Now let us
deal with $\cos\widehat{(\overrightarrow{v^x},\overrightarrow{r^x_m})}$ by making the
following definitions:
\begin{align*}
\theta & = \widehat{(\overrightarrow{v^0},\overrightarrow{v^1})} \\
\theta_m & = \widehat{(\overrightarrow{r^0_m},\overrightarrow{r^1_m})} \\
\Delta_m & = \theta-\theta_m \\
\omega_m & = \widehat{(\overrightarrow{r^0_m}+\overrightarrow{r^1_m},\overrightarrow{v^0}+\overrightarrow{
v^1})}
\end{align*}
$\omega_m$ is the angle between the bisector of
$(\overrightarrow{r^0_m},\overrightarrow{r^1_m})$ and that of
$(\overrightarrow{v^0},\overrightarrow{v^1})$. Given that we want to
minimize $D_m$ in terms of $U_m$ we can safely assume
$\overrightarrow{r^0_m},\overrightarrow{r^1_m},\overrightarrow{v^0},\overrightarrow{v^1}$
to be coplanar. Thus $D_m$ may now be rewritten in terms of those angles as well as $p_m$
and $q_m$:
\begin{align*}
D_m & =\frac{p_m+q_m - p_m \cos(\Delta_m - \omega_m) - q_m
\cos(\Delta_m + \omega_m)}{2} \nonumber
\end{align*}
In this equation the values of $p_m$, $q_m$ and $\Delta_m$ are fully
determined by $\epm$, as described in (\ref{pm}),(\ref{qm}),(\ref{four
scalar product}). $\omega_m$ on the
other hand solely depends on $U_m$: it can be chosen at will by
rotation in the Bloch Sphere. We now show how Eve must tune
$\omega_m$ so as to minimize $D_m$.
\begin{align*}
 \frac{\partial D_m}{\partial \omega_m}=  0 
 \Rightarrow p_m \sin(\Delta_m - \omega_m) - q_m \sin(\Delta_m + \omega_m) = 0
\nonumber
\end{align*}
The minimum occurs at:
\begin{align*}
&\omega_m  =  \arcsin \Big(\frac{p_m-q_m}{\sqrt{p_m^2+q_m^2+2p_m q_m\cos(2\Delta_m)}}\Big) \nonumber\\
& \mbox{which yields, after simplification:}\nonumber\\
&D_m  =\frac{p_m+q_m-\sqrt{p_m^2+q_m^2+2p_m q_m\cos(2\Delta_m)}}{2}
\nonumber
\end{align*}
\subsection{The Tradeoff}
How many elements should Eve's measurement contain? Levitin has
proved that there exists a two-element measurement $\{M_m\}_{m=0,1}$
which maximizes Eve's Information Gain \cite{Levitin}. While this was never
formally shown to be the case for the measurements which optimize
the Information Gain versus Disturbance Tradeoff, there is strong
numerical evidence in support of this assumption \cite{Fuchs}. Suppose this
is the case and let $\epm_{\mu}$ denote the ${\mu}^{th}$ coordinate of
$\epm$. Using the constraint equation (\ref{sum to identity}) we have:
\begin{align}
\delta \ul{\ep_0}_{\mu} & =- \delta \ul{\ep_1}_{\mu}\
\quad \mbox{and thus:} \nonumber\\
\forall f \quad \frac{\partial f}{\partial\ul{\ep_0}_{\mu}} & =
-\frac{\partial f}{\partial \ul{\ep_1}_{\mu}} \label{take and pay}
\end{align}
Optimizing the Tradeoff implies finding a stationary point for the
Disturbance while keeping the Information Gain fixed. We need
to find $\ul{\ep_0}$ such that
\begin{align*}
&\sum_{\mu} \frac{\partial D}{\partial\ul{\ep_0}_{\mu}} \delta \ul{\ep_0}_{\mu} =  0 \\
\end{align*}
where the variations $\delta \ul{\ep_0}_{\mu}$ are subject to the
additional constraint:
\begin{align*}
&\sum_{\mu} \frac{\partial I}{\partial\ul{\ep_0}_{\mu} } \delta \ul{\ep_0}_{\mu}
=  0 \\
\end{align*}
Using equation (\ref{take and pay}) and $D=D_0+D_1$ and $I=I_0+I_1$ this gives:
\begin{align}
\sum_{\mu} \frac{\partial D_0}{\partial \ul{\ep_0}_{\mu}} \delta \ul{\ep_0}_{\mu} &= 
\sum_{\mu} \frac{\partial D_1}{\partial\ul{\ep_1}_{\mu} } \delta \ul{\ep_0}_{\mu}
\label{constant D}\\
\mbox{subject to} \quad \sum_{\mu} \frac{\partial I_0}{\partial \ul{\ep_0}_{\mu}} \delta \ul{\ep_0}_{\mu} & = 
\sum_{\mu} \frac{\partial I_1}{\partial \ul{\ep_1}_{\mu}} \delta \ul{\ep_0}_{\mu}
\label{constant I}
\end{align}
Guided by the geometrical picture of the scenario one may consider
the following attack: 
\begin{align*}
\underline{\ep_0} & =  [\begin{array}{cccc} 1 \;&\beta \; \, \: \! &0 &0 \end{array}] \\
\underline{\ep_1} & =  [\begin{array}{cccc} 1 &-\beta &0 &0
\end{array}]
\end{align*}
The fact that this is indeed a solution follows from its obvious symmetries:
\begin{align}
\textrm{For} \; \mu\neq 2 \quad\frac{\partial
D_0}{\partial\ul{\ep_0}_{\mu} } & =  \frac{\partial D_1}{\partial
\ul{\ep_1}_{\mu}}\;\;\textrm{and for}\; \mu= 2 \quad\frac{\partial D_0}{\partial \ul{\ep_0}_{\mu} } = - \frac{\partial
D_1}{\partial\ul{\ep_1}_{\mu} } \label{mirror I}\\
\textrm{For} \; \mu\neq 2 \quad\frac{\partial
I_0}{\partial\ul{\ep_0}_{\mu} } & =  \frac{\partial
I_1}{\partial\ul{\ep_1}_{\mu} } \;\;\textrm{and for} \; \mu= 2 \quad\frac{\partial I_0}{\partial\ul{\ep_0}_{\mu} } = -
\frac{\partial I_1}{\partial \ul{\ep_1}_{\mu}} \label{mirror D}
\end{align}
Substituting (\ref{mirror D}) in  the constant Information constraint
(\ref{constant I}), we get $\delta \ul{\ep_0}_2 = 0$. Using this fact together with
equation (\ref{mirror I}) it becomes clear that condition
(\ref{constant D}) is
fulfilled. Thus $\ul{\ep_0}$ is a stationary point. We may now proceed to compute the values of the
Disturbance and the Information Gain under this family of optimal
attacks. First by making a few additional observations:
\begin{align*}
p_0&=q_1= p \\
p_1&=q_0= q \\
D_0&=D_1= D/2 \\
I_0&=I_1= I/2 \\
D& = \frac{1}{2}-\sqrt{p^2+q^2+2p q+\cos(2\Delta_m)} \\
I & =  1+2p \log(2p)+2q\log(2q)
\end{align*}
and second by plugging in the relations (\ref{three
scalar product}), (\ref{v0 coord})-(\ref{qm}), we reproduce the
exact content of Fuchs and Peres' formulae:
\begin{align*}
D& =
\frac{1}{2}-\frac{1}{2}\sqrt{1+(c^2-c^4)(\beta^2-2+2\sqrt{1-\beta^2})} \nonumber\\
I & =
\frac{1}{2}((1+\beta c)\log(1+\beta c)+(1-\beta c)\log(1-\beta
c)) \nonumber
\end{align*}

\section{Conclusion}

In this paper we considered a linear embedding taking $d \times d$ positive
hermitian matrices into vectors of $d^2$ real entries, $\phi: 
\rho\mapsto \ul \rho = (\trace(\rho \tau_{\mu}))_{\mu}$. 
It is a well-known fact that the most general evolution a density
matrix $\rho$ may undergo is
a generalised measurement $\{M_m\}_m=\{U_m\sqrt{E_m}\}_m$, where the
polar decomposition was applied. In order to represent $M_m$'s
per-outcome effect upon
the real vectors we defined $\psi:A \mapsto \phi \circ Ad_A \circ
\phi^{-1}$ and
showed that $\psi(U_m)$ is a real orthogonal transform while    
$\psi(\sqrt{E_m})$ turns out to be a real positive matrix. Thus the
geometrical effect of a generalized measurement can be viewed in terms
of real transformations only. \\
Such a nice correspondence suggests quantum mechanics could be
expressed elegantly over the real numbers in this manner, quite
differently from its formulation
in terms of real Jordan algebras \cite{Jordan}. However we first need to gain more geometrical intuition about the
set of real vectors $\phi(\textrm{Herm}^+_d(\mathbb{C}))$, and the sets of allowed
 orthogonal and positive transforms. For now we know that $\phi(\textrm{Herm}^+_d(\mathbb{C}))$ is
a subcone of the future-light-cone $\Gamma= \{(\lambda_{\mu})\in
\mathbb{R}^{d^2} / \sum_{i=1}^{d^2-1}\lambda_i^2 \leq
(d-1)\lambda_0^2, \lambda_0 \geq 0 \} $. \\
One of the advantages of defining $\phi$ upon $\textrm{Herm}^+_d(\mathbb{C})$ instead
of the restricted set of density matices is that  $E_m=M_mM_m^{\dagger}$ can be
visualized. In order to characterize its effects we derived rather compact and powerful
formulae for the qubit case, such as the one giving the scalar product of the
post-measurement states:
\begin{align*}
\frac{1}{4}[2(\ul{E_m}.\ul{\rho^0})(\ul{E_m}.\ul{\rho^1})-(\eta_{\mu
\mu'} \ul{E_m}_{\mu}\ul{E_m}_{\mu'})(\eta_{\nu
\nu'}\ul{\rho^0}_{\nu}\ul{\rho^1}_{\nu'})]
\end{align*}
By looking at such expressions it becomes apparent that Minkowskian products have
a crucial role to play in our framework, and even more so as we showed
that pure quantum states correspond to light-like vectors (i.e. they sit
on the boundary of $\Gamma$), even in dimensions greater than $2$. It
seems interesting to notice that BPS states of supersymmetric theories
can also be thought of as
lying on the boundary of a cone of positive operators \cite{BPS}, and
that their stability is related to that fact. Somehow our setting
seems to single out generalized pure states in a more natural way than
merely characterizing them as unit rank elements of the boundary of
$\textrm{Herm}_d^+(\mathbb{C})$. The reminiscence of special relativity must be
investigated further; this will be a subject for future work.\\
 Pauli matrices together with special relativistic considerations have
already brought some fruitful results to quantum information theory. This
is the case for instance in \cite{Gisin}, where some limits of quantum
cloning are derived by using the no-signalling condition. Armed with
the present representation one should be able tackle more of these
difficult quantum information theoretical problems. Already in this
paper we recovered Fuchs and Peres' information gain versus disturbance formula simply and
geometrically. In the future we should be able to extend our analysis
to the case of two non-equiprobable states. Some highly symmetric
$n$-states scenarios may well cease to be out of reach.

\section{Acknowlegments}
C.E.P would like to thank Gary Gibbons for motivating discussions on
convex cones, EPSRC, the DAMTP, and the Cambridge European and Isaac Newton Trusts for financial support.
P.J.A  would like to thank Louis Salvail for lengthy discussions on
methods for information gain versus disturbance tradeoff, Anuj
Dawar for his patient listening, EPSRC, Marconi, the Cambridge
European  and Isaac Newton Trusts for financial support.

\end{document}